\documentclass[12pt]{article}
\usepackage{graphicx}
\usepackage{amsmath, amssymb}
\usepackage{hfstyle}
\newtheorem{proposition}{Proposition}
\newtheorem{conjecture}{Conjecture}

\newcommand{\NN}{{\mathbf{N}}}

\begin{document}
\author{Henryk Fuk\'s
      \oneaddress{
         Department of Mathematics\\
         Brock University\\
         St. Catharines, Ontario  L2S 3A1, Canada\\
        \texttt{hfuks@brocku.ca}}}
\title{Probabilistic cellular automata with conserved quantities}
\Abstract{We demonstrate that the concept of a conservation law
can be naturally extended from deterministic to probabilistic
cellular automata (PCA) rules. The local function for conservative
PCA must satisfy conditions analogous to conservation conditions
for deterministic cellular automata. Conservation condition for
PCA can  also be written in the form of a current conservation
law. For deterministic  nearest-neighbour CA the current can be
computed exactly. Local structure approximation can partially
predict the equilibrium current for non-deterministic cases. For
linear segments of the fundamental diagram it actually produces
exact results.} \maketitle
\section{Introduction}
Cellular automata (CA) are dynamical systems characterized by
discreteness in space and time. In general, they can be viewed as
cells in a regular lattice updated synchronously according to a
local interaction rule, where the state of each cell is restricted
to a finite set of allowed values.

As in any other dynamical system, conservation laws play an
important role in CA. Additive invariants in one-dimensional CA
have been studied by Hattori and Takesue \cite{Hattori91}. They
obtained conditions which guarantee the existence of additive
conserved quantities, and produced a table of additive invariants
for Wolfram's elementary CA rules. The simplest of additive
invariants, namely the number of active sites (``active'' meaning
non-zero), plays especially important role in CA dynamics. CA
possessing such invariant, to be called ``conservative CA'', can
be viewed as a system of interacting particles, as described in
\cite{paper10}. In a finite system, the flux or current of
particles in equilibrium depends only on their density, which is
invariant. The graph of the current as a function of density
characterizes many features of the flow, and is therefore called
the fundamental diagram. Fundamental diagrams of conservative CA
were investigated in \cite{paper8,paper12}. For majority of
conservative CA rules, fundamental diagrams are piecewise-linear,
usually possessing one or more ``sharp corners'' or singularities.
Although the shapes of fundamental diagrams vary, there is a
strong evidence of the universal behavior at singularities, as
reported in \cite{paper19}.

Conservative CA appear in various applications, and some special
cases have been studied extensively. Rule 184, which is a discrete
version of the totally asymmetric exclusion process, is a prime
example of such special case
\cite{Krug88,Nagatani95,Nagel96,Belitsky98,Sipper96,paper11,
NishinariT98,BelitskyKNS01,Blank03}. Although much is known about
this particular rule, dynamics of other conservative rules
exhibits many features which are not fully understood, and more
general results are just starting to appear. For example, M.
Pivato \cite {Pivato02} recently studied conservation laws in CA
in a very general setting, deriving both theoretical consequences
and practical tests for conservation laws, and provided a method
for constructing all one-dimensional CA exhibiting a given
conservation law. Another recent work \cite{Moreira03}
investigates universality and decidability of conservative CA.
Unfortunately, there exists no general result explaining the shape
of fundamental diagrams for conservative CA's, in spite of a
remarkable progress reported recently for a special class of rules
\cite{KujalaL02}.

In this paper, we will introduce a natural extension of the
conservation condition to include probabilistic cellular automata
(PCA). Deterministic conservative rules will then become a special
case of conservative PCA. For the nearest-neighbour case, this
allows to parameterize all conservative PCA by a set of three
parameters. In the three-dimensional parameter space,
nearest-neighbour conservative PCA are represented by a
polyhedron, with deterministic rules located at its vertices. We
will then show that the general shape of the fundamental diagram
for nearest-neighbour PCA can be partially explained by a
mean-field type approximation.

\section{Probabilistic boolean CA}

In what follows, we will assume that the dynamics takes place on a
one-dimensional lattice of length $L$ with periodic boundary
conditions. Let $s_i(k)$ denote the state of the lattice site $i$
at time $k$, where $i \in \{0,1,\ldots,L-1\}$, $k\in \NN$. All
operations on spatial indices $i$ are assumed to be modulo $L$. We
will further assume that $s_i(k)\in \{0,1\}$, and we will say that
the site $i$ is occupied (empty) at time $k$ if $s_i(k)=1$
($s_i(k)=0$). We will also define
$\mathbf{s}(k)=\{s_0(k),s_1(k),\ldots, s_{L-1}(k)\}$.

In a probabilistic cellular automaton, lattice sites
simultaneously change states form $0$ to $1$ or from $1$ to $0$
with probabilities depending on states of local neighbours. A
common method for defining PCA is to specify a set of local
transition probabilities. For example, in order to define a
nearest-neighbour PCA one has to specify the probability
$w(s_i(k+1))| s_{i-1}(k),s_i(k),s_{i+1}(k))$ that the site
$s_i(k)$ with nearest neighbors $s_{i-1}(k),s_{i+1}(k)$ changes
its state to $s_i(k+1)$ in a single time step. For the sake of
illustration, consider a recently investigated PCA rule
\cite{paper20}, where empty sites become occupied with a
probability proportional to the number of occupied sites in the
neighborhood, and occupied sites become empty with a probability
proportional to the number of empty sites in the neighborhood. The
following set of transition probabilities defines the
aforementioned PCA rule:
\begin{alignat}{2} \label{difPCA}
w(1|0,0,0)&=0    & \qquad  w(1|0,0,1)&=p     \nonumber\\
w(1|0,1,0)&=1-2p & \qquad  w(1|0,1,1)&=1-p   \nonumber\\
w(1|1,0,0)&=p    & \qquad  w(1|1,0,1)&=2p    \nonumber\\
w(1|1,1,0)&=1-p  & \qquad  w(1|1,1,1)&=1,
\end{alignat}
where $p\in [0,1/2]$. The remaining eight transition probabilities
can be obtained using $w(0|a,b,c)=1-w(1|a,b,c)$ for
$a,b,c\in\{0,1\}$.

We will now proceed to construct a general definition of PCA, with
arbitrary neighbourhood size. Let $n$ be a positive integer, and
let $C=\{c_1,c_2,\ldots,c_n\}$ be a set of integers such that
$c_{i+1}=c_i+1$ for all $i=1\ldots n-1$. The set
$\{s_{i+c_1}(k),s_{i+c_2}(k),\ldots,s_{i+c_n}(k)\}$ will be called
a \emph{neighbourhood} of the site $s_i(k)$. We will assume that
the neighbourhood always includes the site $s_i(k)$, i.e.,
$c_1\leq 0$.

Consider now a set of independent Boolean random variables
$X_{i,\mathbf{v}}$, where $i=0,1,\ldots L-1$ and $\mathbf{v} \in
\{0,1\}^n$. Probability that the random variable
$X_{i,\mathbf{v}}$ takes the value $a \in\{0,1\}$ will be denoted
by $w(a|\mathbf{v})$,
\begin{equation}\label{defw}
Pr(X_{i,\mathbf{v}}=a)=w(a|\mathbf{v}).
\end{equation}
Obviously, $w(1|\mathbf{v})+w(0|\mathbf{v})=1$ for all $\mathbf{v}
\in \{0,1\}^n$. The update rule for PCA is defined by
\begin{equation}\label{defprobca}
  s_i(k+1)=X_{i,\{s_{i+c_1}(k),s_{i+c_2}(k),\ldots,s_{i+c_n}(k)\}}.
\end{equation}
Note that $\{\mathbf{s}(k) : k=0,1,2,\ldots\}$ is a Markov
stochastic process, and its states are binary sequences
$\mathbf{s}(k) \in \{0,1\}^L$. The probability of transition  from
$\mathbf{x}\in \{0,1\}^L$ to $\mathbf{y}\in \{0,1\}^L$ in one time
step is given by
\begin{equation}\label{markovprop}
    Pr(\mathbf{y} | \mathbf{x})=\prod_{i=0}^{L-1}
    w(y_i|\{x_{i+c_1},x_{i+c_2},\ldots,x_{i+c_n}\}).
\end{equation}

To illustrate the above formalism, let us consider again the rule
(\ref{difPCA}) defined at the beginning of this section. For this
rule, we have $C=\{-1,0,1\}$, and each lattice site $i$ is
associated with eight random variables
 $X_{i,000}$,
 $X_{i,001}$,
 $X_{i,010}$,
 $X_{i,011}$,
 $X_{i,100}$,
 $X_{i,101}$,
 $X_{i,110}$, and
 $X_{i,111}$.
Probability distributions of these r.v. are determined by
(\ref{difPCA}). For instance, $w(1|1,0,0)=p$, and therefore we
have $Pr\big(X_{i,100}=1\big)=p$, and
$Pr\big(X_{i,100}=0\big)=1-p$. Similarly, $w(1|0,0,0)=0$, hence
$Pr\big(X_{i,000}=1\big)=0$ and $Pr\big(X_{i,000}=0\big)=1$,
meaning that $X_{i,000}$ is in this case a deterministic variable.

\section{Conservative rules}
 If for any initial distribution $\mu$
\begin{equation}\label{consdef}
    E_\mu \left(\sum_{i=0}^{L-1}s_i(k)\right) = E_\mu \left(\sum_{i=0}^{L-1}s_i(k+1)\right)
\end{equation}
for all $k=0,1,\ldots$, then the PCA defined by (\ref{defprobca})
will be called conservative. The expectation value of the sum
$\sum_{i=0}^{L-1}s_i(k)$ can be interpreted as the expected number
of sites occupied by ``particles'', therefore the condition
(\ref{consdef}) requires that the expected number of ``particles''
in the system is constant.

Since the expected value of the random variable $X_{i,\mathbf{v}}$
depends only on the vector $\mathbf{v}$, we will now define
function $f: \{0,1\}^n \rightarrow [0,1]$ as
\begin{equation}\label{defoff}
    f(x_1,x_2,\ldots , x_n)=E
    \left(X_{i,\{x_1,x_2,\ldots,x_n\}}\right).
\end{equation}

\begin{proposition}
The probabilistic CA rule defined by (\ref{defprobca}) is
conservative if and only if
\begin{equation}\label{consf}
    \sum_{i=0}^{L-1} f(x_i,x_{i+1},\ldots, x_{i+n-1})=
    \sum_{i=0}^{L-1} x_i
\end{equation}
for all $\{x_0,x_1, \ldots , x_{L-1}\} \in \{0,1\}^L$.
\end{proposition}
\textbf{Proof.} To prove that (\ref{consdef}) implies
(\ref{consf}), it is enough to choose deterministic initial
configuration, $s_0(0)=x_0$, $s_1(0)=x_1$, \ldots,
$s_{L-1}(0)=x_{L-1}$ for any $\{x_0,x_1, \ldots , x_{L-1}\} \in
\{0,1\}^L$. Then
\begin{equation}
E_\mu \left(\sum_{i=0}^{L-1}s_i(0)\right) =\sum_{i=0}^{L-1} x_i,
\end{equation}
and using (\ref{defprobca})
\begin{multline}
 E_\mu \left(\sum_{i=0}^{L-1}s_i(1)\right)
 =\sum_{i=0}^{L-1} E_\mu
\left(X_{i,\{s_{i+c_1}(0),s_{i+c_2}(0),\ldots,s_{i+c_n}(0)\}}\right)\\
=\sum_{i=0}^{L-1} E_\mu
\left(X_{i,\{x_{i+c_1},x_{i+c_2},\ldots,x_{i+c_n}\}}\right)
=\sum_{i=0}^{L-1} f(x_{i+c_1},x_{i+c_2},\ldots,x_{i+c_n}\})\\
=\sum_{j=0}^{L-1} f(x_j,x_{j+1},\ldots, x_{j+n-1}),
\end{multline}
where $j=i+c_1$.  Condition (\ref{consdef}) requires that $E_\mu
\left(\sum_{i=0}^{L-1}s_i(0)\right)$ and $ E_\mu
\left(\sum_{i=0}^{L-1}s_i(1)\right)$ must be equal, hence
(\ref{consf}) follows.

In order to prove that  (\ref{consf}) implies (\ref{consdef}), we
will use the definition of the expectation value
\begin{equation}
 E_\mu \left(\sum_{i=0}^{L-1}s_i(k+1)\right)
 =\sum_{\mathbf{x} \in \{0,1\}^L} Pr\big(\mathbf{s}(k+1)=\mathbf{x}\big) \sum_{i=0}^{L-1}
 x_i,
\end{equation}
where $\mathbf{x}=\{x_0,x_1,\ldots,x_{L-1}\}$. Probability
$Pr\big(\mathbf{s}(k+1)=\mathbf{x}\big)$ can be written as
\begin{equation}
    Pr\big(\mathbf{s}(k+1)=\mathbf{x}\big)=\sum_{\mathbf{y} \in \{0,1\}^L}
    Pr(\mathbf{x} | \mathbf{y})
    Pr\big(\mathbf{s}(k)=\mathbf{y}\big)
\end{equation}
where $Pr(\mathbf{x} | \mathbf{y})$ denotes the probability of
transition from $\mathbf{y}$ to $\mathbf{x}$ in a single time
step. Combining the above two equations and changing the order of
summation  we obtain
\begin{multline}
E_\mu \left(\sum_{i=0}^{L-1}s_i(k+1)\right) \\
 =\sum_{\mathbf{x}
\in \{0,1\}^L} \sum_{\mathbf{y} \in \{0,1\}^L}
Pr\big(\mathbf{s}(k)=\mathbf{y}\big)  Pr(\mathbf{x}|
\mathbf{y})\sum_{i=0}^{L-1} x_i \\
=  \sum_{\mathbf{y} \in \{0,1\}^L}
Pr\big(\mathbf{s}(k)=\mathbf{y}\big) \sum_{i=0}^{L-1}
\sum_{\mathbf{x} \in \{0,1\}^L} Pr(\mathbf{x}| \mathbf{y})\, x_i.
\end{multline}
Note that the last sum, $\sum_{\mathbf{x} \in \{0,1\}^L}
Pr(\mathbf{x}| \mathbf{y})\, x_i$, is simply equal to the expected
value of $x_i$ given the previous state of the system
$\mathbf{y}$. This expected value must be equal to
\begin{equation}
    E
    \left(X_{i,\{y_{i+c_1},y_{i+c_2},\ldots,y_{i+c_n}\}}\right)=
f(y_{i+c_1},y_{i+c_2},\ldots,y_{i+c_n}),
\end{equation}
and as a consequence we have
\begin{multline}
E_\mu \left(\sum_{i=0}^{L-1}s_i(k+1)\right) \\
=\sum_{\mathbf{y} \in \{0,1\}^L}
Pr\big(\mathbf{s}(k)=\mathbf{y}\big) \sum_{i=0}^{L-1}
f(y_{i+c_1},y_{i+c_2},\ldots,y_{i+c_n})\\
=\sum_{\mathbf{y} \in \{0,1\}^L}
Pr\big(\mathbf{s}(k)=\mathbf{y}\big) \sum_{i=0}^{L-1} y_i\\
=E_\mu \left(\sum_{i=0}^{L-1}s_i(k)\right),
\end{multline}
precisely what we wanted to show.\hfill$\square$

Following \cite{Hattori91,paper12}, it is possible to obtain a
simple condition which a probabilistic conservative CA rule must
satisfy.
\begin{proposition} \label{Hattoricond}
A probabilistic CA rule $f$ is number-conserving if, and only if,
for all $\{x_1,x_2,\ldots,x_n\} \in \{0,1\}^n$, it satisfies
\begin{align}
f(x_1,x_2,\ldots,x_n) = x_1 + \sum_{k=1}^{n-1}\big(
&f(\underbrace{0,0,\ldots,0}_k,x_2,x_3,\ldots,x_{n-k+1})\notag\\
-&f(\underbrace{0,0,\ldots,0}_k,x_1,x_2,\ldots,x_{n-k})\big).
\label{NScond}
\end{align}
\end{proposition}
\textbf{Proof.} To prove that Condition (\ref{NScond}) is
necessary, we will first note that $f(0,0,\ldots,0)=0$, which is a
direct consequence of the condition (\ref{consf}) applied to the
configuration consisting of all zeros, $\{x_0,x_1, \ldots ,
x_{L-1}\}=\{0,0,\ldots, 0\}$. Consider now a configuration of
length $L\ge 2n-1$ which is the concatenation of a sequence
$\{x_1,x_2,\ldots,x_n\}$ and a sequence of $L-n$ zeros. If $f$ is
conservative,  such configuration must satisfy condition
(\ref{consf}), hence
\begin{align}
f(0,0,\ldots,0,x_1)& + f(0,0,\ldots,0,x_1,x_2) + \cdots\notag\\
&\qquad + f(x_1,x_2,\ldots,x_n) + f(x_2,x_3,\ldots,x_n,0)+\cdots\notag\\
&\qquad + f(x_n,0,\ldots,0) = x_1+x_2+\cdots+x_n, \label{x1neq0}
\end{align}
where all the terms of the form $f(0,0,\ldots,0)$, which are equal
to zero, have not been written.

Replacing $x_1$ by 0 in (\ref{x1neq0}) gives
\begin{align}
f(0,0,\ldots,0,x_2)& + \cdots + f(0,x_2,\ldots,x_n)\notag\\
& + f(x_2,x_3,\ldots,x_n,0) + \cdots + f(x_n,0,\ldots,0)\notag\\
& = x_2+\cdots+x_n. \label{x1eq0}
\end{align}
Subtracting (\ref{x1eq0}) from (\ref{x1neq0}) yields
(\ref{NScond}).

To prove that condition (\ref{NScond}) is sufficient, we can apply
it to each site $x_i$ of a configuration
$\{x_0,x_1,\ldots,x_{L-1}\}$,
\begin{multline}
f(x_{i+1},x_{i+2},\ldots,x_{i+n}) \\= x_{i+1} +
\sum_{k=1}^{n-1}\big(
f(\underbrace{0,0,\ldots,0}_k,x_{i+2},x_{i+3},\ldots,x_{i+n-k+1})\\
-f(\underbrace{0,0,\ldots,0}_k,x_{i+1},x_{i+2},\ldots,x_{i+n-k})\big).
\end{multline}
Obviously, when the above is summed over $i\in
\{0,1,\ldots,L-1\}$, all the right-hand side terms except the
first cancel, and we obtain (\ref{consf}). \hfill$\square$

Local mappings $f$ for conservative PCA exhibit some interesting
symmetries. We have already demonstrated that $f(0,0,\ldots,0)=0$,
and one can easily show that similar property is true when we
replace zeros by ones, i.e.,  $f(1,1,\ldots,1)=1$. The following
proposition shows yet another symmetry.
\begin{proposition}
If a mapping $f:\{0,1\}^n \rightarrow [0,1]$ represents
conservative PCA rule, then
\begin{equation}\label{balanced}
    \sum_{x_1,x_2, \ldots, x_n \in\{ 0,1\}} f(x_1,x_2, \ldots, x_n) =
2^{n-1}.
\end{equation}
We will say that $f$ satisfying the above condition is
$1$-balanced.
\end{proposition}

\textbf{Proof.} Consider a configuration
$\mathbf{t}=\{t_0,t_1,\ldots,t_{L-1} \}\in \{0,1\}^L$, where
$L=2^n$. Now, let us construct a set of sequences of length $n$,
$A=\{\mathbf{b}^{(j)}\}_{j=0}^{L-1}$ such that
$\mathbf{b}^{(j)}=\{t_j,t_{j+1},\ldots, t_{j+n-1}\}$. Superscript
$(j)$ denotes here just a consecutive number of the sequence
$\mathbf{b}^{(j)}$ in the set $A$. Recall that all operations on
subscript indices are taken modulo $L$.

Assume that we can find $\mathbf{t}$ such that all sequences in
$A$ are different. This means that each possible sequence of
length $n$ occurs in $A$ once and only once, and therefore
\begin{equation} \label{t1}
\displaystyle \sum_{x_1,x_2, \ldots, x_n \in\{ 0,1\}}  f(x_1,x_2,
\ldots, x_n) = \sum_{i=0}^{L-1} f(t_i,t_{i+1},\ldots,
t_{i+n-1})=N_{1},
\end{equation}
where $N_{1}$ is the number of 1's in the sequence $\mathbf{t}$.

On the other hand, $\bar{\mathbf{t}}\in \{0,1\}^L$, which is
obtained from $\mathbf{t}$ by replacing all zeros by ones and {\em
vice versa}, must also have the same property as $\mathbf{t}$,
i.e.,
\begin{equation} \label{t2}
\displaystyle \sum_{x_1,x_2, \ldots, x_n \in\{ 0,1\}}  f(x_1,x_2,
\ldots, x_n) = \sum_{i=0}^{L-1} f(\bar{t}_i,\bar{t}_{i+1},\ldots,
\bar{t}_{i+n-1})=N_{0},
\end{equation}
where $N_{0}$ is the number of 0's in $\mathbf{t}$. Comparing
(\ref{t1}) and (\ref{t2}) we obtain $N_0=N_1=2^n/2=2^{n-1}$,
exactly as required.

The only problem left is to show that, indeed, for any $n>0$, we
can construct the sequence $\mathbf{t}$ of length $2^n$ (with
periodic boundary conditions) such that all subsequences of length
$n$ occurring in $t$ are different (and therefore, constitute a
set of all possible sequences of length $n$). For example, for
$n=3$, $\mathbf{t}=11101000$ is the required sequence. One can see
that sequences of length $3$ occurring in $\mathbf{t}$, $111$,
$110$, $101$, $010$, $100$, $000$, $001$, and $011$, are all
possible binary sequences of length $3$.

For a general $n$, the required $\mathbf{t}$ is equivalent to a
hamiltonian cycle in the de Bruijn graph  \cite{DeBruij46} of
dimension $n$ (or an eulerian cycle in the de Bruijn graph of
dimension $n-1$). It can be demonstrated that such a cycle always
exists (more precisely, for a given $n$, there exist exactly
$2^{2^{n-1}-n}$ of such cycles -- see, for example, review article
\cite{Ralston}). \hfill$\square$

\section{Example: nearest neighbour conservative PCA}
We will now consider the case of nearest-neighbour PCA, for which
the neighbourhood is defined by $C=\{-1,0,1\}$. The most general
PCA of this type is defined by eight parameters $ w(1|abc)\in
[0,1]$ for $a,b,c\in\{0,1\}$:
\begin{alignat}{2}
 f(0, 0, 0) &=w(1|000) ,& \qquad  f(1, 0, 0) &=w(1|100) , \nonumber\\
 f(0, 0, 1) &=w(1|001) ,& \qquad  f(1, 0, 1) &=w(1|101) , \nonumber \\
 f(0, 1, 0) &=w(1|010) ,& \qquad  f(1, 1, 0) &=w(1|110) , \nonumber \\
 f(0, 1, 1) &=w(1|011) ,& \qquad  f(1, 1, 1) &=w(1|111). \label{generalw}
\end{alignat}
  Remaining eight probabilities can be determined
 by using consistency condition $w(0|abc)+w(1|abc)=1$.
Proposition \ref{Hattoricond} imposes extra conditions on $f$,
\begin{equation}\label{nnconscond}
    f(x_1,x_2,x_3)=x_1 + f(0,x_2,x_3)+f(0,0,x_2)
    -f(0,x_1,x_2)-f(0,0,x_1),
\end{equation}
for all $x_1,x_2,x_3 \in \{0,1\}$. Writing these conditions
explicitly, one obtains eight equations, among them only five are
independent, implying that there are three free parameters in the
solution. The general solution parameterized by $\alpha$, $\beta$
and $\gamma$ is given by
\begin{alignat}{2}
 f(0, 0, 0) &=w(1|000)= 0   ,           & \quad  f(1, 0, 0) &=w(1|100)= \alpha, \nonumber\\
 f(0, 0, 1) &=w(1|001)= \beta ,         & \quad  f(1, 0, 1) &=w(1|101)= \alpha+\beta, \nonumber \\
 f(0, 1, 0) &=w(1|010)= 1-\alpha-\beta , & \quad  f(1, 1, 0) &=w(1|110)= 1 -\beta+\gamma, \nonumber \\
 f(0, 1, 1) &=w(1|011)= 1-\alpha-\gamma,& \quad  f(1, 1, 1) &=w(1|111)= 1, \label{nnconsgen}
\end{alignat}
where the three free parameters $\alpha,\beta,\gamma$ satisfy
conditions
\begin{equation} \label{polyconditions}
\alpha \geq 0, \quad \beta \geq 0, \quad \alpha+\beta \leq 1,
\quad \alpha+\gamma \leq 1, \quad \beta - \gamma \leq 1,
\end{equation}
so that all probabilities $w(1|abc)$ remain in the interval
$[0,1]$.
  One can easily show that the following single
expression for $f(x_1,x_2,x_3)$  is equivalent to
(\ref{nnconsgen}):
\begin{equation}\label{genfclosedform}
 f(x_1,x_2,x_3) = \gamma (x_1 x_2 - x_2 x_3)
  +\alpha x_1 + (1-\alpha-\beta) x_2 + \beta x_3,
\end{equation}
where $x_1,x_2,x_3 \in \{0,1\}$.

The set of points $(\alpha,\beta,\gamma)$ in 3D space satisfying
(\ref{polyconditions}) forms a polyhedron which has the shape of a
pyramid with rhomboidal base and triangular sides, as shown in
Figure~\ref{elcons}. There are five possible choices of parameters
$\alpha,\beta,\gamma$ leading to purely deterministic rules (where
all probabilities $w(1|abc)$ are either zero or one), and they are
listed  in Table~\ref{elcons}. These deterministic rules
correspond to five vertices of the pyramid of Figure~\ref{elcons}.

\begin{table}
  \centering
  \begin{tabular}{|c|c|c|c|c|}
  \hline
  $\alpha$ & $\beta$ & $\gamma$ & Rule number & $f(x_1,x_2,x_3)$ \\
  \hline
 1 & 0 & 0 & 240 &  $x_1$\\
 1 & 0 & -1 & 184 &  $x_1-x_1 x_2+x_2 x_3$\\
 0 & 0 & 0 & 204 &  $x_2$\\
 0 & 1 & 1 & 226  & $x_3 + x_1 x_2 - x_2 x_3$\\
 0 & 1 & 0 & 170 &  $x_3$\\
  \hline
\end{tabular}
  \caption{Five deterministic elementary conservative CA.}\label{elcons}
\end{table}
\begin{figure}
\begin{center}
\includegraphics[scale=0.5]{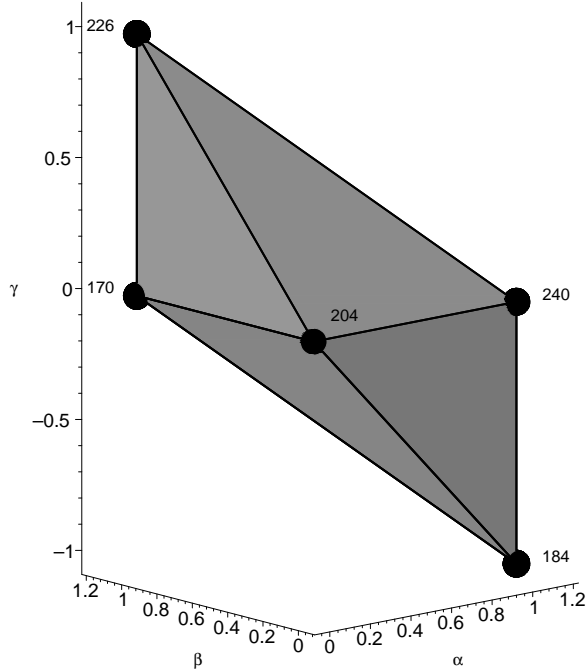}
\end{center}
\caption{Visual representation of the set of all conservative PCA
rules in the parameter space $(\alpha,\beta,\gamma)$.}
\label{piramidy}
\end{figure}

\section{Current conservation and fundamental diagrams}
In \cite{Hattori91}, Hattori and Takesue demonstrated that
existence of a conserved quantity in a deterministic CA is
equivalent to a discrete version of a standard current
conservation law $\partial \rho/\partial t=-\partial j/\partial
x$. In what follows, we will show that a similar current
conservation law holds for conservative PCA.

We will first observe that the sum on the right hand side of the
condition (\ref{Hattoricond}) can be split into two parts, one
depending on $x_2,\ldots,x_{n}$, and another which depends on
$x_1,x_2,\ldots,x_{n-1}$. We will therefore define the function of
$n-1$ arguments
\begin{equation}\label{curdef}
    J(x_1,x_2,\ldots,x_{n-1})=
    - \sum_{k=1}^{n-1}
f(\underbrace{0,0,\ldots,0}_k,x_1,x_2,\ldots,x_{n-k}) +
\sum_{j=1}^{-c_1} x_j,
\end{equation}
With this definition, it is straightforward to show that
\begin{proposition}
A probabilistic CA rule $f$ is number-conserving if, and only if,
for all $\{x_1,x_2,\ldots,x_n\} \in \{0,1\}^n$, it satisfies
\begin{equation}\label{curcons}
    f(x_1,x_2,\ldots,x_n)- x_{-c_1+1}=J(x_1,x_2,\ldots,x_{n-1})
    -J(x_2,x_3,\ldots,x_n).
\end{equation}
\end{proposition}
Applying this to all lattice sites we obtain
\begin{multline} \label{allsitescons}
    f(s_{i+c_1},s_{i+c_2},\ldots,s_{i+c_n})-
    s_{i}=J(s_{i+c_1},s_{i+c_2},\ldots,s_{i+c_{n-1}})\\
    -J(s_{i+c_2},s_{i+c_3},\ldots,s_{i+c_{n-1}}),
\end{multline}
where we dropped $k$ dependence for clarity.

In order to understand the full meaning of the above equation, let
us recall a general formulation of a conservation law in a
continuous, one dimensional physical system. Let $\rho(x,t)$
denote the density of some material at point $x$ and time $t$, and
let $j(x,t)$ be the current (flux) of this material at point $x$
and time $t$. A conservation law states that the rate of change of
the total amount of material contained in a fixed domain is equal
to the flux of that material across the surface of the domain. The
differential form of this condition can be written as
\begin{equation}\label{concons}
    \frac{\partial \rho}{\partial t}=-\frac{\partial j}{\partial
    x}.
\end{equation}

Interpreting $s_i(k)$ as the density, the left hand side of
(\ref{allsitescons}) is simply the expected change of density in a
single time step, so that (\ref{allsitescons}) is an obvious
discrete analog of the current conservation law (\ref{concons})
with $J$ playing the role of the current. We can pursue this
analogy even further. For the partial differential equation
(\ref{concons}) one often assumes a functional relation between
the current $j$ and the density $\rho$, and we will do the same
for conservative PCA: we will ask how the ``average'' current $J$
depends on the ``average'' density in the stationary state,
assuming that we start from a disordered initial condition.

To be more precise, let us now assume that the initial
distribution $\mu$ is a Bernoulli distribution, i.e., at $k=0$,
all sites $s_i(k)$ are independently occupied with probability $p$
or empty with probability $1-p$, where $p\in [0,1]$. Let us define
$\rho(i,k)=E_\mu (s_i(k))$. Since the initial distribution is
$i$-independent, we expect that $\rho(i,k)$ also does not depend
on $i$, and we will therefore define $\rho(k)=\rho(i,k)$.
Furthermore, for conservative PCA, $\rho(k)$ is $k$-independent,
so we define $\rho=\rho(k)$. We will refer to $\rho$ as the
density of occupied sites. The expected value of the current
$J(s_{i+c_1}(k),s_{i+c_2}(k),\ldots,s_{i+c_{n-1}}(k))$ will also
be $i$-independent, so we can define the expected current as
\begin{equation} \label{defexpcurrent}
j(k,\rho)=E_\mu
\big(J(s_{i+c_1}(k),s_{i+c_2}(k),\ldots,s_{i+c_{n-1}}(k))\big).
\end{equation}

The graph of the  equilibrium current $j(\rho)=\lim_{k \rightarrow
\infty} j(k,\rho)$ versus the density $\rho$ is known as the
fundamental diagram. It has been numerically demonstrated
\cite{paper8} that for conservative deterministic CA the
fundamental diagram usually develops a singularity as
$k\rightarrow \infty$, meaning that $j(\rho)$ is not everywhere
differentiable function of $\rho$.

\section{Current and fundamental diagrams for nearest-neighbour
PCA}

For nearest-neighbour conservative PCA with the function $f$ given
by eq. (\ref{genfclosedform}), the current becomes
\begin{equation}\label{currentnearestneig}
    J(x_1,x_2)= x_1 - f(0,x_1,x_2) - f(0,0,x_1)=
    \gamma x_1 x_2+\alpha x_1 - \beta x_2,
\end{equation}
and the conservation condition (\ref{curcons}) becomes
\begin{equation}\label{curconsnn}
    f(x_1,x_2,x_3)- x_{2}=J(x_1,x_2)-J(x_2,x_3).
\end{equation}
For the five nearest-neighbour deterministic rules, the current
$J(x_1,x_2)$ assumes particularly simple form. For rule 184, the
expression (\ref{currentnearestneig}) reduces to $J(x_1,x_2)= x_1
(1-x_2)$, which means that the current $J(s_i(k),s_{i+1}(k))$ is
non-zero only if $s_i(k)=1$ and $s_i(k)=0$. This agrees with the
interpretation of rule 184 as a system of interacting particles,
in which a particle located at site $i$ moves to site $i+1$ only
if the site $i+1$ is empty. Rule $226$ is similar, except that
particles move in the opposite direction than in the rule 184. For
rules $240$, $170$ and $204$, the current becomes, respectively,
$J(x_1,x_2)=x_1$, $J(x_1,x_2)=-x_2$, and $J(x_1,x_2)=0$. This
means that all particles move to the right (rule 170), to the left
(rule 240), or stay in the same place (rule 204).

It turns out that the simplicity of the expression for
$J(x_1,x_2)$ for deterministic nearest-neighbour rules makes it
possible to compute the expected current $j(k,\rho)$ exactly. For
rule $204$, we obviously have $j(k,\rho)=0$. For rules $240$ and
$170$ the expected current is just a linear function of $\rho$,
and we have $j(k,\rho)=\rho$ for rule $240$ and $j(k,\rho)=-\rho$
for rule $170$.

For rule 184 (and its generalizations), one can also compute the
expected current $j(k,\rho)$ exactly, as done in \cite{paper11},
\begin{equation} \label{flowfor184}
j(k,\rho) =1-\rho- \sum_{i=1}^{k+1} \frac{i}{k+1}
  \binom{2k+2}{k+1-i}
  \rho^{k+1-i} (1-\rho)^{k+1+i},
\end{equation}
and one can prove that
\begin{equation}
j(\rho)=\lim_{k \rightarrow \infty}j(k,\rho)= -\left|\rho -
\frac{1}{2}\right|+\frac{1}{2}.
\end{equation}
This means that the graph of $j(\rho)$ versus $\rho$ has a
singularity at $\rho=1/2$, as shown in  Figure~\ref{diagplots}a.
Rule $226$ exhibits similar behavior, except that the direction of
the current is reversed.
\renewcommand{\arraystretch}{1.5}
\begin{table}
  \centering
  \begin{tabular}{|c|c|c|c|}
  \hline
  Rule nr & $J(x_1,x_2)$ & $j(k,\rho)$ &$j(\rho)$\\
  \hline
 240 & $x_1$        &   $\rho$ &$\rho$\\ 
 184 & $x_1(1-x_2)$ &   $\displaystyle 1-\rho- \sum_{i=1}^{k+1} \frac{i}{k+1} \binom{2k+2}{k+1-i} \rho^{k+1-i}
 (1-\rho)^{k+1+i}$& $\displaystyle -\left|\rho -
\frac{1}{2}\right|+\frac{1}{2}$\\ 
 204 & $0$          &   $0$ & $0$\\ 
 226 & $(1-x_1)x_2$ &   $\displaystyle-1+\rho+ \sum_{i=1}^{k+1} \frac{i}{k+1} \binom{2k+2}{k+1-i}
 \rho^{k+1-i}(1-\rho)^{k+1+i}$& $\displaystyle \left|\rho -
\frac{1}{2}\right|-\frac{1}{2}$ \\ 
 170 & $-x_2$       &   $-\rho$ & $-\rho$\\
  \hline
\end{tabular}
  \caption{Expected current $j(k,\rho)$ and the
  expected equilibrium current $j(\rho)$ for
  five deterministic elementary conservative CA.}\label{conscurrents}
\end{table}
Expressions for $j(k,\rho)$ and $j(\rho)$ for all five
deterministic nearest-neighbour rules are summarized in
Table~\ref{conscurrents}. Unfortunately, these results cannot be
easily generalized to compute expected currents for probabilistic
rules. Even the equilibrium current $j(\rho)$ for the general rule
given by eq. (\ref{genfclosedform}) does not seem to be
analytically tractable, not to mention the time-dependent current
$j(k,\rho)$. In the following section, we will use mean-field type
technique to obtain approximate equilibrium current for
probabilistic rules.

\section{Mean Field Current}
For nearest-neighbour rules, we can write the following expression
for the average current, using equations (\ref{defexpcurrent}) and
(\ref{currentnearestneig}),
\begin{equation}
j(k)=\gamma E_{\mu}\big(s_{i-1}(k) s_i(k) \big)+ (\alpha-\beta)
\rho,
\end{equation}
where we used the fact that $E_{\mu}\big(s_i(k) \big)=\rho$. We
will use abbreviated notation for probabilities $Pr(s_{i-1}(k),
s_i(k) =a_1 a_2)$, denoting them simply by $P_k(a_1 a_2)$, where
$a_1,a_2 \in \{0,1\}$. $P_k(11)$ is in this notation the
probability that the pair $(s_{i-1}(k), s_i(k))$ is at time step
$k$ in the state $(1,1)$, and since
\begin{equation}
E_{\mu}\big(s_{i-1}(k) s_i(k) \big)= Pr\big((s_{i-1}(k), s_i(k))
=(1,1)\big),
\end{equation}
we obtain
\begin{equation} \label{jasprob}
j(k)=\gamma P_k(1,1)+ (\alpha-\beta) \rho.
\end{equation}

By (\ref{markovprop}), probabilities $P_k(a_1 a_2)$ must satisfy
the following relationship
\begin{equation}\label{recur}
P_{k+1}(a_1 a_2)=\sum_{b_1,b_2,b_3,b_4 \in \{0,1\}} w(a_1
a_2|b_1b_2b_3b_4 ) P_k(b_1 b_2 b_3 b_4),
\end{equation}
where $w(a_1 a_2|b_1b_2b_3b_4 )=w(a_1|b_1b_2b_3) w( a_2|b_2b_3b_4
)$, and the values of transition probabilities $w$ are given by
(\ref{generalw}). As before, $P_k(b_1 b_2 b_3 b_4)$ is the
probability that the sequence of four consecutive sites at time
$k$ assumes value $(b_1, b_2, b_3, b_4)$.

The equation (\ref{recur}) gives two-site probabilities at time
$k+1$ in terms of four-site probabilities at time $k$. In order to
make it useful, we have to eliminate four-site probabilities from
this equation. This can be done only approximately, using Bayesian
extension formula also known as local structure theory
\cite{gutowitz87a},
\begin{equation} \label{locstr}
    P_k(b_1 b_2 b_3 b_4) \approx
    \frac{P_k(b_1 b_2) P_k(b_2 b_3) P_k(b_3 b_4)}{P_k(b_2)
    P_k(b_3)},
\end{equation}
where $P_k(1)=\rho$, $P_k(0)=1-\rho$.

This approximation transforms (\ref{recur}) into a set of
recurrence relations for pair probabilities $P_k(a_1 a_2)$. Fixed
point of this set corresponds to the stationary state, and the
knowledge of the  fixed point value of $P_k(11)$  allows us to
find $j(\rho)$ by (\ref{jasprob}).

Although computation of the aforementioned fixed point is too long
and tedious to be attempted by hand, it can be carried out by a
computer algebra software. The resulting stationary value of the
current is shown below (we used ``mf'' subscript to indicate
mean-field nature of the approximation (\ref{locstr})):
\begin{equation} \label{localstructurecurrent}
j_{\mathrm{mf}}(\rho)= (\alpha-\beta+\gamma) \rho - \frac{A \gamma
\rho (\rho-1) }{2(\gamma \rho - \gamma + \beta )(-\gamma \rho +
\gamma + \alpha )}
\end{equation}
where
\begin{align}\nonumber
A=(\alpha-\beta+\gamma)(\alpha-\beta+\gamma - \gamma \rho) -\alpha
-\beta +\alpha \beta \\
-{\rm sgn}(\gamma) \sqrt{c_0+c_1 \rho+c_2 \rho^2},
\end{align}
\begin{align} \nonumber
c_{{0}}={\beta}^{4}-2 \left( 2 \gamma+1+ \alpha \right) {\beta
}^{3}+ \left( -5 {\alpha}^{2}+4 \gamma-4 \gamma \alpha+1 +2
{\gamma}^{2} \right) {\beta}^{2}\\ \nonumber+ 2\left(\alpha  -
{\alpha}^{3}- {\gamma}^{2}+3 \alpha {\gamma}^{2}+2 \gamma {\alpha
}^{2}\right) \beta-2 \alpha
{\gamma}^{2}-2 {\alpha} ^{3}\\
+4 {\alpha}^{3}\gamma+{\gamma}^{4}+{\alpha}^{2}-4 \gamma
{\alpha}^{2}+{\alpha}^{4}+2 {\gamma}^{2}{\alpha}^{2},
\end{align}
\begin{align}\nonumber
c_{{1}}=-2 {\gamma}^{4}+ 2\left(  \alpha- \beta \right) {{
 \gamma}}^{3}+ 2 \left(
  \alpha+\beta - {\alpha}^{2}- {\beta}^{2}-7 \alpha  \beta
\right) {\gamma}^{2} \\
+2 \left({\beta}^{3} - {\alpha}^{3}+ {\alpha}^{2}- {\beta}^{2} +2
{\beta}^{2}\alpha- 2 {\alpha}^{2}\beta \right) \gamma
\end{align}
and
\begin{equation}
c_{{2}}={\gamma}^{2} \left({\alpha}^{2}+{\beta}^{2}+\gamma^{2} -2
\gamma \alpha+6 \alpha \beta+2 \gamma \beta
 \right).
\end{equation}
The above approximation does a remarkably good job in predicting
$j(\rho)$ for deterministic rules. In fact, it yields all five
expressions for the equilibrium current reported in the last
column of Table~\ref{conscurrents}.

\begin{figure}
\begin{center}
\includegraphics[scale=0.9]{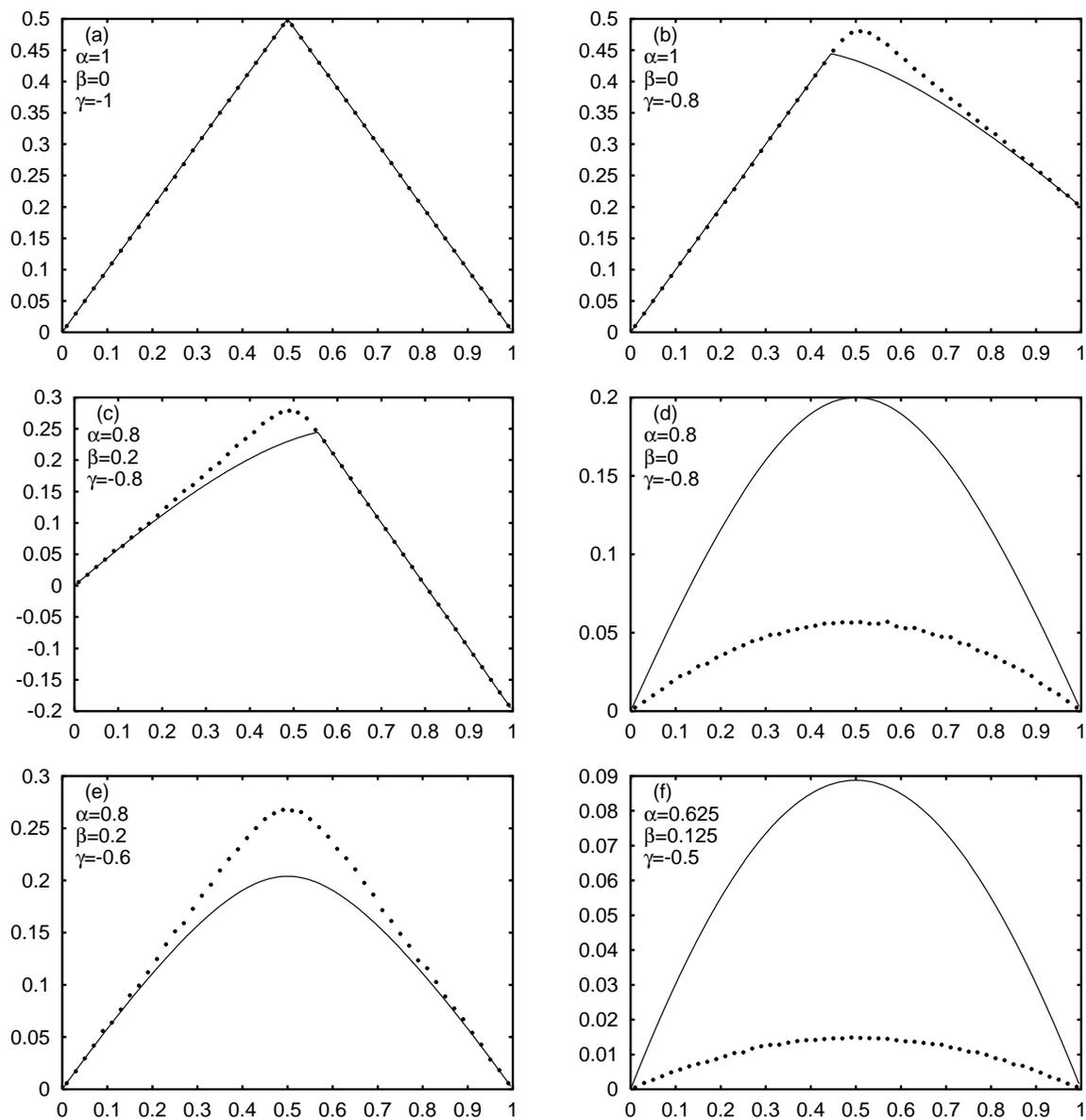}
\end{center}
\caption{Fundamental diagrams of selected nearest-neighbour
conservative PCA rules. Horizontal axis represents the density
$\rho$, while the vertical axis represents the stationary current
$j(\rho)$. Solid dots represent simulation results, and the
continuous lines are graphs of $j_{MF}(\rho)$ vs. $\rho$.}
\label{diagplots}
\end{figure}

For non-deterministic cases, the approximation does not work
equally well. In order to evaluate its accuracy, we performed
computer simulations in which the current $j(k,\rho)$ was recorded
for large values of $k$, thus approximating the equilibrium
current $j(\rho)$. Results are reported in Figure~\ref{diagplots},
where we show fundamental diagrams for a selection of five
nearest-neighbour rules. Dotted lines represent simulation
results, while continuous lines represent the approximate current
given by (\ref{localstructurecurrent}). Values of parameters
$\alpha$, $\beta$ and $\gamma$ are shown in the upper left corner
of each diagram. Only one diagram, namely Figure~\ref{diagplots}a,
corresponds to a deterministic rule (rule 184). All other diagrams
represent non-deterministic rules.

Diagram (b) correspond to the rule located on the edge of the
polyhedron of Figure~\ref{elcons} joining rules 184 and 240, close
to the vertex of rule $184$. Similarly, diagram (c) represents the
rule located on the edge joining rules 184 and 170, again close to
the vertex of rule $184$. Remarkably, it appears that the local
structure theory predicts correctly the shape of the linear part
of the fundamental diagram, while the nonlinear part is only
approximately correct. It also predicts existence of the
singularity in the fundamental diagram, while numerical
simulations indicate that no such singularity exists.

When we consider a rule located on the edge joining rule 184 and
rule 204, as shown in diagram (d), no part of the fundamental
diagram is predicted correctly by the local structure theory
(LST). The shape of the LST diagram, however, is not far from the
observed shape. One can note similar phenomenon for a rule located
on the base of the pyramid of Figure~\ref{elcons} (diagram e) or
in the interior of the pyramid (diagram f).

The above evidence suggest some conjectures about fundamental
diagrams of conservative PCA. First of all, singularity of the
fundamental diagram appears to be a  purely deterministic
phenomenon. The fundamental diagram has a singularity (at
$\rho=1/2$) only for rules 184 and 226. For all other rules, the
graph of $j(\rho)$ has no ``sharp corners''. Recall that the
general probabilistic CA rule is defined in terms of probabilities
$w(a|\mathbf{v})$. If all of these probabilities belong to the set
$\{0,1\}$, such a rule will be called \emph{deterministic}. All
other PCA rules will be called \emph{truly probabilistic}.
\begin{conjecture}
The equilibrium current $j(\rho)$ for any truly probabilistic
conservative CA rule is a differentiable function of $\rho$ for
$\rho \in (0,1)$.
\end{conjecture}
Secondly, linear parts of fundamental diagrams can be computed
exactly by using local structure approximation. This has been
observed in \cite{paper8} for deterministic rules, and appears to
be true for probabilistic rules as well.
\begin{conjecture}
If the equilibrium current $j(\rho)$ of a probabilistic
conservative CA rule is a linear function of $\rho$ in some
interval $(a,b)\subseteq (0,1)$, then the local structure
approximation for $j(\rho)$ in this interval is exact.
\end{conjecture}

\section{Conclusion}

We have demonstrated that the concept of a conservation law can be
naturally extended from deterministic to probabilistic cellular
automata rules. Local function $f$ for conservative PCA must
satisfy conditions analogous to conservation conditions for
deterministic CA. We also proved that the local function must be
1-balanced. For nearest-neighbour rules, conservation conditions
reduce the number of free parameters defining the rule to three,
making it possible to visualize all nearest-neighbour rules as
points of a polyhedron in 3D space, with deterministic rules
located at vertices of the polyhedron.

Conservation condition for PCA can be also written as a current
conservation law. For deterministic  nearest-neighbour rules the
current can be computed exactly, but for probabilistic rules it
appears to be a much more difficult problem. Nevertheless, local
structure approximation can partially predict the equilibrium
current. For linear segments of the fundamental diagram it
actually produces exact results.

Although our conjectures regarding  fundamental diagrams are
supported by numerical evidence for nearest-neighbour rules,
obviously more evidence is needed, especially for larger
neighbourhood sizes. Preliminary evidence indicates that both
conjectures hold for four-input rules and five-input rules, and
these findings will be reported elsewhere.

Another interesting question is the issue of representation. In
\cite{paper10}, it has been demonstrated that conservative
deterministic CA can be viewed as systems of interacting
particles, or, in other words, that conservative CA rules can be
defined using ``motion representation'' (as rules defining motion
of individual particles). Similar concept has been introduced in
\cite{Pivato02} as a ``displacement representation''. As pointed
out in \cite{MatsukidairaN03}, the motion or displacement
representation is analogous to Lagrange description of the fluid
flow in hydrodynamics, in which we observe a particle and follow
its trajectory. Standard definition of the CA rule (as in eq.
\ref{defprobca}), on the other hand, resembles Euler description
of the flow, where the flow is observed at a fixed point in space
and the dependent variable represents the amplitude of a field at
that point. For deterministic conservative CA, both Langrange and
Euler descriptions are possible. It seems that it should be
possible to construct Lagrange description for probabilistic
conservative CA as well, although it is not clear how to
accomplish this task. The main problem is that the particles are
not strictly conserved in PCA, only their expected number is.
Since a given  particle may disappear at any time, it is not
possible to choose one particle and follow its trajectory
indefinitely. Some averaging procedure will clearly be needed
here.

 \vskip 1cm
 \noindent \textbf{Acknowledgements:} The author
acknowledges financial support from NSERC (Natural Sciences and
Engineering Research Council of Canada) in the form of the
Discovery Grant.

\providecommand{\href}[2]{#2}\begingroup\raggedright\endgroup

\end{document}